\documentclass[aps,prl,preprint,superscriptaddress,showpacs]{revtex4}
\usepackage[dvips]{graphicx}

\begin{document}

\title{ Excitonic Polaron States and Fano Interference in Phonon-Assisted Photoluminescence of Bound Excitons in ZnO }

\author{Shi-Jie Xu}
\email{sjxu@hkucc.hku.hk}
\address{Department of Physics and HKU-CAS Joint Laboratory on
New Materials, The University of Hong Kong, Pokfulam Road, Hong Kong, China}

\author{Shi-Jie Xiong}
\email{sjxiong@nju.edu.cn}
\address{Department of Physics and HKU-CAS Joint Laboratory on
New Materials, The University of Hong Kong, Pokfulam Road, Hong Kong, China}
\address{National Laboratory of Solid State Microstructures and
Department of Physics, Nanjing University, Nanjing 210093, China}

\author{Shen-Lei Shi}
\address{Department of Physics and HKU-CAS Joint Laboratory on
New Materials, The University of Hong Kong, Pokfulam Road, Hong Kong, China}
\begin{abstract}

We report a photoluminescence observation of robust excitonic polarons due to strong coupling of exciton and longitudinal optical (LO) phonon as well as Fano-type interference in high quality ZnO crystal. At low enough temperatures, the strong coupling of excitons and LO phonons leads to not only traditional Stokes lines (SLs) but also up to second-order anti-Stokes lines (ASLs) besides the zero-phonon line (ZPL). The SLs and ASLs are found to be not mirror symmetric with respect to the ZPL, strongly suggesting that they are from different coupling states of exciton and phonons. It is more interesting that a new group of peaks, including a ZPL and several SLs, are observed. The observations can be explained with a newly developed theory in which this group is attributed to the ground excitonic polaron state and the other group is from the excited polaron states with LO phonon components partially decaying into environal phonon modes. Besides these spectral features showing the quasiparticle properties of exciton-phonon coupling system, the first-order SL is found to exhibit characteristic Fano lineshape, caused by quantum interference between the LO components of excitonic polarons and the environal phonons. These findings lead to a new insight into fundamental effects of exciton-phonon interaction. 
\end{abstract}


\maketitle


As a II-VI wide-gap polar semiconductor with the wurtzite structure, zinc oxide (ZnO) is of great technological importance. Since the demonstration of lasing from its microcrystalline film \cite{1}, ZnO has attracted renewed interests in recent years \cite{2,3,4,5,6,7,8,9,10,11,12}. ZnO has some unique properties such as large exciton binding energy ($\sim$ 60 meV) and strong exciton-LO phonon interaction \cite{12}. While most of the recent efforts have been devoted to understanding such issues as defects \cite{2,3,4,11}, surface polarity \cite{6,8}, and p-type doping \cite{5,10} etc., few reports were presented to investigate the exciton-phonon interaction in ZnO, despite that this interaction was recognized to be fundamentally important for its optoelectronic properties from both experiments \cite{12} and theories \cite{13,14}. Strong exciton-phonon interaction may lead to the existence of new exciton-phonon bound states \cite{13}, which causes striking phonon-assisted exciton transitions, referred to as phonon sidebands or replicas, in absorption and emission spectrum. For example, the energy separations of the phonon sidebands with respect to the zero-phonon line in emission and absorption spectra can be quite different. For emission, the energy separation is close to the standard characteristic energy of LO phonons. For absorption, however, the energy separation is significantly smaller. In ZnO, the characteristic energy of the $\text{A}_1-\text{LO}$ phonon at zone center is about $\sim$ 72 meV ($\sim$ 574 $cm^{-1}$) \cite{15}. The binding energy of shallow excitons in ZnO may be comparable to the characteristic energy of LO phonons, leading to interesting effects of resonant coupling between exciton and LO phonons \cite{16}. A similar resonant coupling in semiconductor quantum dots has attracted an increasing interest \cite{17,18,19,20}. Such resonant mixing between exciton and phonons has been argued to form new quasiparticles, namely excitonic polarons \cite{20} and to produce a new class of nonadiabatic lines in phonon-assisted luminescence of exciton \cite{19}, which are difficult to be interpreted by the use of the previous adiabatic theory of multiphonon transitions in deep centers developed by Huang and Rhys \cite{21,22}. 
    At the same time, another resonance phenomenon-Fano-type quantum interference \cite{23} has been suggested to exist in the coupling exciton-phonon systems of semiconductors \cite{14,24,25,26}. However, to our knowledge, it has not yet been experimentally observed in excitonic luminescence. In this Letter, we report the first luminescence evidence of the existence of excitonic polaron states and phonon-mediated Fano-type quantum interference in high-quality ZnO single crystal. A new theory beyond adiabatic approximation is developed to account for almost all the spectral features observed.
    
      The ZnO samples used in the present study are commercial bulk crystals (Commercial Crystal Laboratories). The 325 nm line of a He-Cd laser was employed to illuminate the Zn-terminated ZnO $(000\bar{1})$ surface or O-terminated ZnO surface at a tilted angle of about $45^{o}$. The low-temperature photoluminescence setup employed in the present work was previously described elsewhere \cite{27}. 
\begin{figure}[h]
\includegraphics[width=9.6cm]{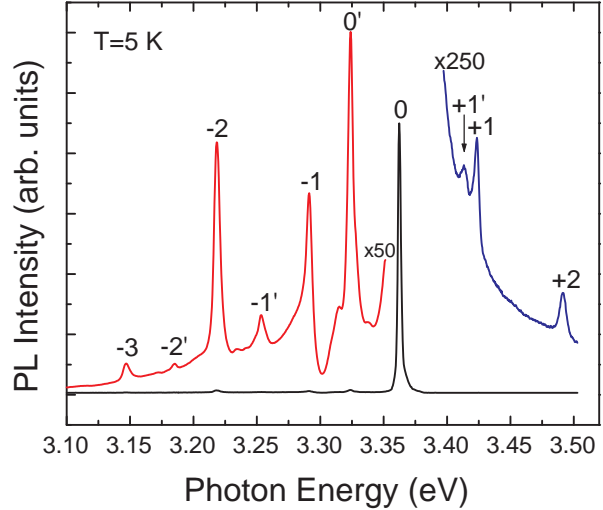}
\caption {Representative photoluminescence spectrum of the exciton-phonon coupling system in ZnO measured at 5 K. Various phonon-assisted transitions including anti-Stokes lines and Fano-type interference lineshape were clearly observed. }
\label{fig1}
\end{figure}

Figure 1 shows a representative low-temperature emission spectrum of the ZnO sample with Zn-terminated surface. The strongest peak at $\sim$ 3.361 eV (369 nm), which is marked as 0, is conventionally identified to be the zero-phonon line (ZPL) of the bound excitons (neutral-donor-bound excitons \cite{28}, usually called $\text{I}_2$ line in literature). The direct band gap of ZnO at low temperature is 3.437 eV \cite{11}. The energy difference between the zero-phonon line and ZnO band gap is thus equal to $(3.437-3.361)\text{eV}=76 \text{meV}$, which is slightly larger than the characteristic energy (72 meV) of the $\text{A}_1-\text{LO}$ at zone center. This provides a physical configuration in which the resonant mixing of excitons and LO phonons occurs. As a result, the nonadiabatic lines should be observed in emission spectra of ZnO excitons. From the emission spectrum in Fig. 1, it is obvious that the first, second, and third phonon Stokes lines (SLs) of the bound excitons, which are denoted with -1, -2, and -3, respectively, are all clearly resolved. These lines represent optical processes in which one, two and three LO phonon creations accompany the corresponding photon generations. The energy separation between any two of the adjacent SLs including the zero-phonon line is $\sim$ 71 meV, which is almost identical to the characteristic energy ($\sim$ 72 meV) of the  $\text{A}_1-\text{LO}$ of ZnO. Besides these ``conventional'' SL lines, the anti-Stokes lines (ASLs, marked as $+1^{'}$, +1 and +2, respectively) are observed at the higher energy side of the ZPL line. Notice that the energy separations between the ASL lines and the ZPL line are noticeably smaller than the corresponding values between the SL lines and the ZPL line. In other words, the ASL and SL lines are not mirror symmetric with respect to the ZPL line. The ASL lines are broader than the SL ones, strongly indicating that the ASL and SL lines are from different coupling exciton-phonon states. Since the thermally excited phonons are negligible at 5 K so that the probability of reabsorbing LO phonons to generate ``hot'' excitons from ``cold'' excitons is negligibly small, the observation of ASL lines directly supports the aforementioned argument. A natural explanation for above spectral features is that the resonant coupling between exciton and phonons produces a number of bound states, i.e., excitonic polaron states \cite{20}. As proved by a newly developed theory (briefly described later) \cite{29}, these bound states contain different exciton and phonon components and give rise to the ASL, ZPL and SL lines. The recorded luminescence observation of the anti-Stokes line was first reported for CdS in 1970 \cite{30}, but the origin was not identified. 
          
      From Fig. 1, it is also seen that there is an additional group of emission lines appearing at the lower energy side of the ZPL at $\sim$ 3.361 eV (marked as 0). Notice that this new group of lines is very similar to the group of lines 0, -1, -2, -3. We thus temporarily mark them $0^{'}$, $-1^{'}$ and $-2^{'}$. One distinct spectral feature in Fig. 1 is the specific asymmetric lineshape of the first phonon SL (marked as -1). Moreover, there exists a clear dip at its higher energy side. It is a typical Fano lineshape \cite{23}. Besides its characteristic Fano lineshape, the intensity of the first-phonon SL is unusually weaker than that of the second phonon SL, which is a scarcely seen phenomenon in cases like ZnO whose Huang-Rhys factor is much less than unity. These anomalies suggest that quantum mechanical interference occurred in the radiative recombination processes of excitonic polarons. 
      
      The observation of phonon-mediated Fano interference phenomenon in luminescence is quite interesting and challenging for the theoretical treatment of the exciton-phonon interaction. First of all, it is recognized that an accurate calculation of the absorption or emission spectrum of a coupled exciton-phonon system in complicated solid-state environment is extremely difficult. Starting from solving Schr\"{o}dinger equation of the exciton-phonon coupling system, we attempt to theoretically reveal these novel spectral features observed in the LO phonon-assisted excitonic PL spectra of ZnO. The Hamiltonian of the exciton-phonon coupling system is given by
\begin{equation}
\label{ham}
 H = \sum_{\bf K} \varepsilon_{\bf K} a^{\dag}_{\bf K} a_{\bf K}
  + \sum_{l,{\bf q}} \hbar \omega_{l,{\bf q}} b^{\dag}_{l,{\bf q}} b_{l,{\bf q}}
  + \sum_{l,{\bf K}, {\bf q}} V_{l,{\bf q}} a^{\dag}_{{\bf K}-{\bf q}} a_{\bf K} \left(
b^{\dag}_{l,{\bf q}} + b_{l,-{\bf q}} \right) ,
\end{equation}
where $a_{\bf K}$ ($a^{\dag}_{\bf K}$) and $b_{l,{\bf q}}$ ($b^{\dag}_{l,{\bf q}}$) are annihilation (creation) operators for exciton with mass-center momentum ${\bf K}$ and  for phonon in the $l$th band with momentum ${\bf q}$, respectively. $\varepsilon_{\bf K}$ and $\omega_{l,{\bf q}}$ characterize the dispersion relations of excitons and phonons. $V_{l,{\bf q}}= \left[ \frac{2\pi e^2 \hbar \omega_{l,{\bf q}}}{Vq^2} \left(\kappa_{\infty}^{-1} - \kappa_{0}^{-1} \right) \right]^{1/2}$ is the Fr\"{o}hlich coupling strength of exciton and LO phonon. Here, $\kappa_{\infty}$, $\kappa_0$, and $V$ are the dielectric constants for infinite and zero frequencies and the volume in a cell, respectively. For localized excitons bound at impurities discussed in the present work, we can neglect its dispersion and let $\varepsilon_{\bf K}\equiv \varepsilon_0$. For Hamiltonian (\ref{ham}), the $m$th eigen-wavefunctions can be written as
\begin{equation}
\label{ent}
  \Psi_m = \sum_{ n_{l,{\bf q}}} c_{m, \{ n_{l,{\bf q}} \}  }
   a^{\dag}\prod_{l,{\bf q}} \frac{ \left( b^{\dag}_{l,{\bf q}} \right)^{n_{l,{\bf q}} }}{\sqrt{n_{l,{\bf q}}!}} | 0 \rangle ,
\end{equation}
where $n_{l,{\bf q}}$ is the number of phonons of mode $(l,{\bf q})$ in the
corresponding component, $|0 \rangle $ is the exciton vacuum, and the
coefficients satisfy the following iteration relations
\[
 \left[ E_m -\left( \varepsilon_0 + \sum_{l,{\bf q}} n_{l,{\bf q}} \hbar \omega_{l,{\bf q}} \right) \right] c_{m, \{ n_{l,{\bf q}} \} }
= \sum_{l',{\bf q}'}  \sqrt{n_{l',{\bf q}'}}
V_{l',{\bf q}'}  c_{m, \{ n_{l,{\bf q}}\} -(l',{\bf q}') }
 \]
 \begin{equation}
\label{eigen}
 +\sum_{l',{\bf q}'} \sqrt{n_{l',{\bf q}'}+1} V_{l',{\bf q}'}
  c_{m, \{ n_{l,{\bf q}}\}+(l',{\bf q}') }.
\end{equation}

    If we further neglect the dispersion of phonons and assume that there is only one LO mode interacting with exciton, we can omit indices $l,{\bf q}$ and rewrite Eq. (\ref{eigen}) as 
\begin{equation}
\label{ss4}
 \left[ E_m -\left( \varepsilon_0 +  n \hbar \omega \right) \right] c_{m,  n  }
=  V_{0} \sqrt{n}\,\,
  c_{m,  n -1 }
 +V_0 \sqrt{n+1} \,\,
  c_{m,  n+1 }.
\end{equation}
Here $V_0$ represents the average exciton-phonon coupling strength. The eigenenergies and corresponding eigenstates of Eq. (\ref{ss4}) can be obtained by cutting the series at a maximum phonon number $n_{\text{max}}$. By this procedure we can obtain $n_{\text{max}}+1$ eigenenergies, and then we increase $n_{\text{max}}$ towards the infinity to yield the converged values. From Eq. (\ref{ss4}), we strictly prove that the well-known Huang-Rhys theory \cite{21} deals with only the ground state (GS) of the excitonic polarons for any coupling strength \cite{29}. The eigenenergy of the GS state is lower than the bare exciton energy by $\Delta_0=\frac{V_0^2}{\hbar\omega}$, and the coefficient of its component containing $n$ LO phonons is $c_{0,n}=\frac{(-1)^nS^{n/2}}{e^{S/2}\sqrt{n!}}$ with $S\equiv \frac{\Delta_0}{\hbar \omega}$ being the Huang-Rhys factor. $|c_{0,n}|^2$ gives the intensities of phonon sidebands predicted by the Huang-Rhys theory. Furthermore, the excited polaron states ($m>0$), which are not handled in the Huang-Rhys theory, can produce ASLs by their components with phonon number $n<m$. The energy positions of these ASLs located at $E_m - n \hbar \omega$. Note that $E_m$ is not exactly equal to $E_0+m\hbar\omega$ due to the exciton-phonon coupling. As a result, the ASLs and SLs are not mirror symmetric about the ZPL, which is consistent with the experimental results in Fig. 1. All these lines form the first set of phonon sidebands marked as 0', -1', -2', +1', and +2' lines in Fig. 1. Besides the LO mode, however, there are a large number of bath modes. Although these modes are not directly interacting with the exciton, they may have a crucial influence on phonon components in the excitonic polaron states. This effect can be taken into consideration with the mixing Hamiltonian
\begin{equation}
  \label{h1}
  H_1=
 \sum_{\lambda} g_{\lambda}
  ( b^{\dag}_0b_{\lambda } +
\text{H.c.}) +\sum_{\lambda}
  \hbar \omega_{\lambda}(b^{\dag}_{\lambda} b_{\lambda } +\frac{1}{2}),
\end{equation}
where $\lambda$ is the mode index of phonons, $g_{\lambda}$ is the coupling strength between the LO and bath modes, and operator $b_0$ is for the LO mode investigated above. Although  all modes are diagonal in a perfect crystal, coupling   may be produced by formation of the exciton polaron states in which the vibrations of the LO phonon components are altered. The dispersion relation $\omega_{\lambda}$ is assumed to be continuous with a constant density of states in an energy range from 0 to $W$. The orthogonal phonon modes under Hamiltonian (\ref{h1}) can be rewritten as
\begin{equation}
  \label{mmode}
   b_l= \frac{ \hbar \omega_l -\hbar \omega}{\sqrt{
   g_{\lambda}^2+(\hbar\omega -\hbar \omega_l)^2}} b_{\lambda}
   +\frac{g_{\lambda}}{\sqrt{g_{\lambda}^2 +(\hbar \omega -\hbar
   \omega_{l})^2}} b_0,
\end{equation}
with frequency $\omega_l \sim \omega_{\lambda}$ as determined by the following equation
\begin{equation}
   \hbar \omega_{l}-\hbar \omega = \sum_{\lambda} \frac{g^2_{\lambda}}{\hbar \omega_l -\hbar \omega_{\lambda}}.
   \end{equation}
The obtained set of modes $\{l\}$ can be used to construct exciton-phonon composites of Eq. (\ref{ent}) and the coupling strength in Eq. (\ref{eigen}) is given by
\begin{equation}
  V_l = \frac{ g_{\lambda}}{ \sqrt{g_{\lambda}^2 +(\hbar \omega
  -\hbar \omega_l)^2}} V_0.
  \end{equation}
\begin{figure}[ht]
\includegraphics[width=9.6cm]{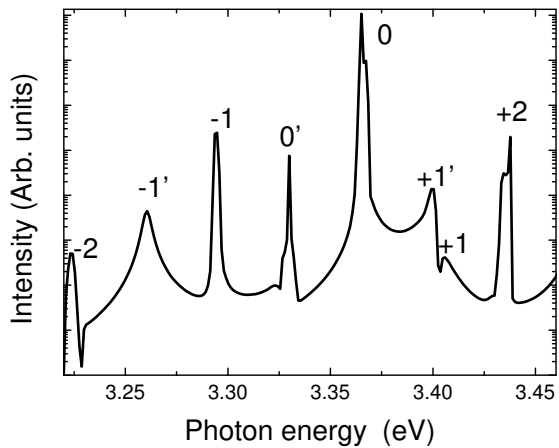}
\caption {Calculated emission spectrum of the exciton-phonon coupling system in ZnO. } \label{fig2}
\end{figure}

  The mixing introduced by $H_1$ is coherent and Fano interference between the discrete LO mode and other continuous modes can take place. When the exciton in $\Psi_m$ is annihilated, photon emission occurs. Simultaneously, the components with phonons in mode $l$ will give rise to an accompanied emission of $l$-mode phonon. Because the $l$ mode is a mixed mode consisting of a LO-mode component and a $\lambda$-mode component as illustrated in Eq. (\ref{mmode}), an interference between these two components will happen, and the probability of corresponding light emission is  proportional to
\begin{equation}
\alpha_l = \frac{(\tilde{\epsilon} + 1)^2 }{\tilde{\epsilon}^2+1},
\end{equation}
with $\tilde{\epsilon}=(\hbar \omega_l-\hbar\omega)/g_{\lambda}$. Such an interference in the optical transitions can lead to a typical Fano lineshape with a unity asymmetric parameter ($q=1$) in the sidebands, especially in the one phonon sideband \cite{23}. Taking the phonon mixing into account, we obtain the PL spectrum
\[
 \rho (\varepsilon)= \sum_{m}  \sum_{\{ n_l \}} |c_{m,\{ n_l \}}|^2
\left( \prod_l \alpha_l^{n_l} \right)\theta(E_{\text{laser}}-E_m)
 \]
 \begin{equation}
  \label{ccc}
 \times \delta(\varepsilon - E_m + \sum_l
n_l \hbar \omega_l ).
\end{equation}
\begin{figure}[ht]
\includegraphics[width=16cm]{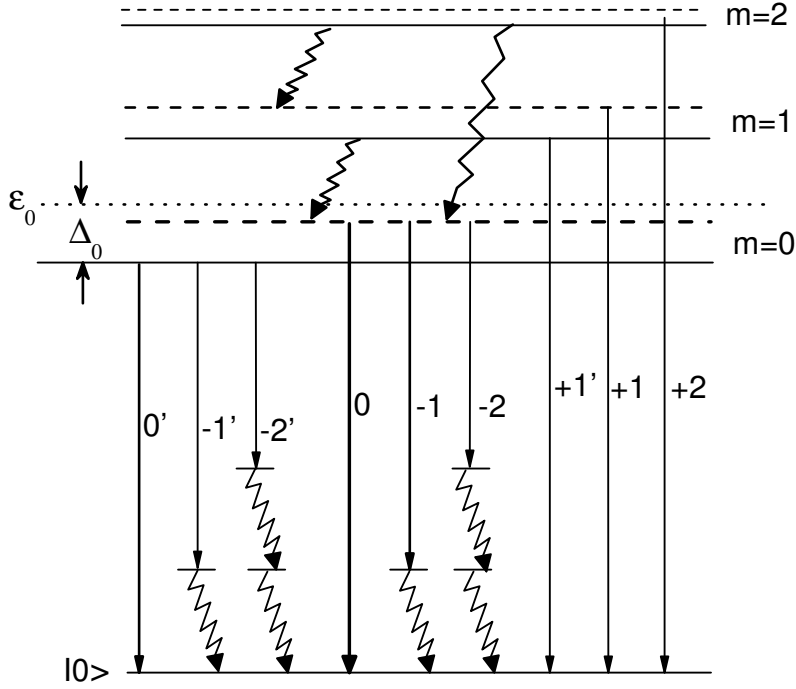}
\caption {Schematic diagram of the excitonic polaron states and relevant optical transitions. 
The solid horizontal lines denote excitonic polaron states $\Psi_m$ comprising exciton and LO phonons. $m=0$ is for the lowest-lying state. The dashed horizontal lines represent states resulting from decaying of LO phonon components into environal phonon modes, and the thick wavy arrows indicate such decaying. The numbers of the solid vertical arrows correspond to the peaks in Figs. 1 and 2, and the thin wavy arrows represent the accompanying emission of LO phonons in corresponding components of the initial excitonic polaron states.
} \label{fig3}
\end{figure}

Using Eq. (\ref{ccc}), we calculate the emission spectrum of ZnO shown in Fig. 2. All the modes $l$ with small coupling strength $V_l$ with the exciton result in the formation of the second set of the phonon sidebands including a ZPL near $\varepsilon_0$ marked as 0, SLs -1, -2, ... and ASLs +1, +2, .... The energy distance between the both ZPLs of the two sets is close to $\Delta_0$ which is determined by the coupling strength $V_0$. Note that the intensity and width of line 0 are much larger than those of 0', because of the huge number of modes with small $V_l$ making the contributions. The theoretical spectrum catches the essential features observed in Fig. 1. The Fano resonance features are partially smeared by the summation in Eq. (\ref{ccc}), but can still be seen for some peaks. Considering the complex of the exciton-phonon coupling system, agreement between theory and experiment is rather satisfactory. According to our calculation, a schematic diagram of the excitonic polaron states and relevant optical transitions is drawn in Fig. 3. All optical transitions observed in the emission spectra can be self-consistently explained using this picture. 
\begin{figure}[ht]
\includegraphics[width=9.6cm]{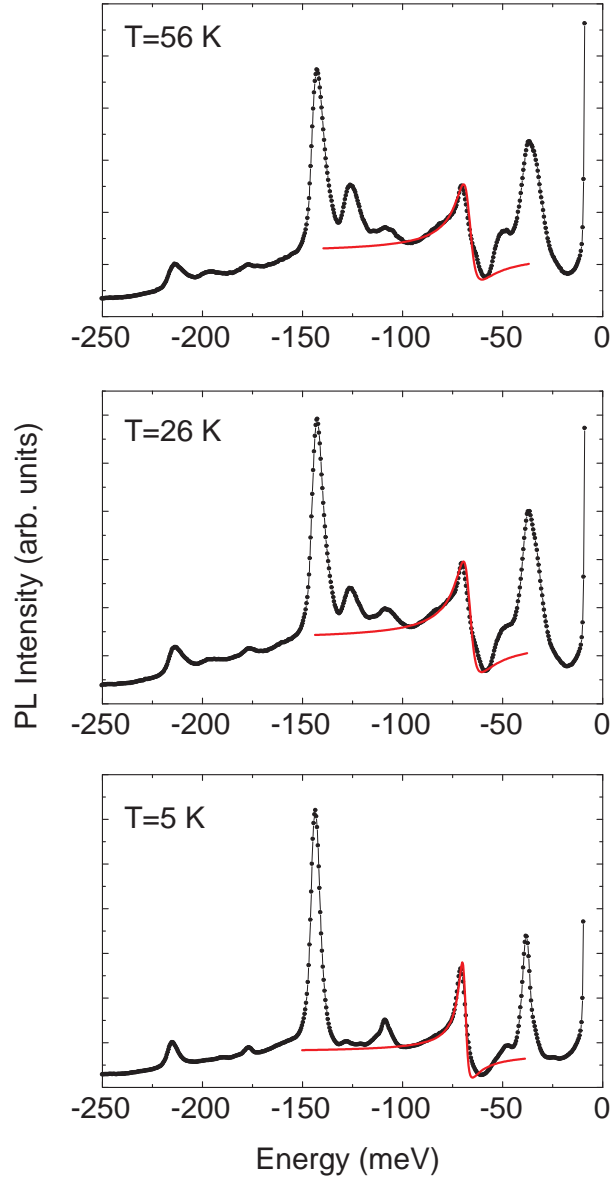}
\caption {Representative emission spectra of the exciton-phonon coupling system at three different temperatures. The energy is measured from the strongest ZPL line. The solid thin lines are the fitting curves using the Fano lineshape function.} \label{fig4}
\end{figure}

    We have used the well-known Fano lineshape function \cite{23} to fit our experimental spectra at different temperatures and the fitting curves at three typical temperatures are plotted in Fig. 4 as the solid lines. The obtained values of the asymmetry parameter $q$ range from 1.43 to 1.85. The theoretical unity value of the parameter is close to them. Another major parameter associated with Fano interference is the resonance width $\Gamma$. It is found that it increases monotonously with increasing temperature in the range of study. This means that the mean lifetime of the excitonic polarons becomes shorter as the temperature increases. From Fig. 1, the line width of the ZPL is $\sim$ 2.5 meV at 5 K. However, the first phonon SL line already broadens to about 4 meV at the same low temperature. The resonance energy $\text{E}_r$  varies within a small range between 67 and 69 meV for the temperature range studied, which is several meV less than the standard LO phonon energy of ZnO. 

      In conclusion, the resonant coupling of bound exciton-LO phonon in high quality ZnO crystal has been investigated with low-temperature photoluminescence. The spectral evidence of the existence of robust excitonic polarons due to the resonant coupling of exciton-phonon was obtained. The phonon mediated Fano-type quantum mechanical interference effect was observed for the first time in the low-temperature luminescence spectra. Together with our theoretical model, the results provide a novel and consistent picture for the exciton-phonon strong coupling system. The work also shows that the Fano-type quantum mechanical interference is a rather general phenomenon in solid state physics.

One of the authors, SJX, would like to acknowledge Prof. S. C. Shen for helpful discussions. It is thankful to Dr. M. H. Xie and Mr. Y. K. Ho for providing samples employed in study. J. Q. Ning participated in early PL measurement. The work was supported in HK by HK RGC CERG Grant (No. HKU 7036/03P), in Nanjing by National Natural Science Foundation of China under contract Nos. 60276005 and 10474033.



\begin{references}

\bibitem{1} P. Zu, Z. K. Tang, G. K. L. Wong, M. Kawasaki, A. Ohtomo, H. Koinuma, and Y. Segawa, Solid State Communi. {\bf 103}, 459 (1997). 
\bibitem{2} D. C. Look, J. W. Hemsky, and J. R. Sizelove, Phys. Rev. Lett. {\bf 82}, 2552 (1999).
\bibitem{3} C. G. Van de Walle, Phys. Rev. Lett. {\bf 85}, 1012 (2000).
\bibitem{4} S. F. J. Cox, E. A. Davis, S. P. Cottrell, P. J. C. King, J. S. Lord, J. M. Gil, H. V. Alberto, R. C. Vil$\tilde{\rm a}$o, J. Piroto Duarte, N. Ayres de Campos, A. Weidinger, R. L. Lichti, and S. J. C. Irvine, Phys. Rev. Lett. {\bf 86}, 2601 (2001).
\bibitem{5} Y. Yan, S. B. Zhang, and S. T. Pantelides, Phys. Rev. Lett. {\bf 86}, 5723 (2001).
\bibitem{6} A. Wander, F. Schedin, P. Steadman, A. Norris, R. McGrath, T. S. Turner, G. Thornton, and N. M. Harrison, Phys. Rev. Lett. {\bf 86}, 3811 (2001).
\bibitem{7} D. M. Hofmann, A. Hofstaetter, F. Leiter, H. Zhou, F. Henecker, B. K. Meyer, S. B. Orlinskii, J. Schmidt, and P. G. Baranov, Phys. Rev. Lett. {\bf 88}, 045504 (2002).
\bibitem{8} O. Dulub, U. Diebold, and G. Kresse, Phys. Rev. Lett. {\bf 90}, 016102 (2003).
\bibitem{9} J. Serrano, F. J. Manj\'{o}n, A. H. Romero, F. Widulle, R. Lauck, and M. Cardona, Phys. Rev. Lett. {\bf 90}, 055510 (2003).
\bibitem{10} L. G. Wang and A. Zunger, Phys. Rev. Lett. {\bf 90}, 256401 (2003).
\bibitem{11} F. Tuomisto, V. Ranki, K. Saarinen, and D. C. Look, Phys. Rev. Lett. {\bf 91}, 205502 (2003).
\bibitem{12} W. Y. Liang and A. D. Yoffe, Phys. Rev. Lett. {\bf 20}, 59 (1968).
\bibitem{13} Y. Toyozawa and J. Hermanson, Phys. Rev. Lett. {\bf 21}, 1637 (1968). 
\bibitem{14} T. Toyozawa, in {\it Optical Processes in Solids}, (Cambridge University Press, 2003), Chap. 10.
\bibitem{15} F. Decremps, J. Pellicer-Porres, A. M. Saitta, J. -C. Chervin, and A. Polian., Phys. Rev. B {\bf 65}, 092101 (2002).
\bibitem{16} J. C. Hermanson, Phys. Rev. B {\bf 2}, 5043 (1970).
\bibitem{17} M. G. Bawendi, W. L. Wilson, L. Rothberg, P. J. Carroll, T. M. Jedju, M. L. Steigerwald, and L. E. Brus, Phys. Rev. Lett. {\bf 65}, 1623 (1990).
\bibitem{18} D. J. Norris, Al. L. Efros, M. Rosen, and M. G. Bawendi, Phys. Rev. B {\bf 53}, 16347 (1996).
\bibitem{19} V. M. Fomin, V. N. Gladilin, J. T. Devreese, E. P. Pokatilov, S. N. Balaban, and S. N. Klimin, Phys. Rev. B {\bf 57}, 2415 (1998).
\bibitem{20} O. Verzelen, R. Ferreira, and G. Bastard, Phys. Rev. Lett. {\bf 88}, 146803 (2002).
\bibitem{21} K. Huang and A. Rhys, Proc. R. Soc. A {\bf 204}, 406 (1950).
\bibitem{22} C. B. Duke and G. D. Mahan, Phys. Rev {\bf 139}, A1965 (1965).
\bibitem{23} U. Fano, Phys. Rev. {\bf 124}, 1866 (1961); U. Fano and J. W. Cooper, Phys. Rev. {\bf 137}, A1364 (1965). 
\bibitem{24} K. P. Jain, Phys. Rev. {\bf 139}, A544 (1965).
\bibitem{25} R. G. Stafford, Phys. Rev. B {\bf 3}, 2729 (1971).
\bibitem{26} J. J. Hopfield, P. J. Dean, and D. G. Thomas, Phys. Rev. {\bf 158}, 748 (1967); P. J. Dean and D. C. Herbert, in {\it Excitons}, ed. by K. Cho, (Springer, Germany, 1979), Chap. 3.   
\bibitem{27} S. J. Xu, W. Liu, and M. F. Li, Appl. Phys. Lett. {\bf 77}, 3376 (2000).
\bibitem{28} R. E. Sherriff, D. C. Reynolds, D. C. Look, B. Jogai, J. E. Hoelscher, T. C. Collins, G. Cantwell, and W. C. Harsch,  J. Appl. Phys. {\bf 88}, 3454 (2000).
\bibitem{29} S. J. Xiong and S. J. Xu, to be published.
\bibitem{30} C. W. Litton, D. C. Reynolds, T. C. Collins, and Y. S. Park, Phys. Rev. Lett. {\bf 25}, 1619 (1970).


\end{references}
\end{document}